\documentclass[notitlepage,
prd,
preprint,
nofootinbib, draft
]{revtex4-1}

\usepackage{etoolbox}

\makeatletter
\patchcmd{\frontmatter@abstract@produce}
  {\vskip200\p@\@plus1fil
   \penalty-200\relax
   \vskip-200\p@\@plus-1fil}
  {}
  {}
  {}
\makeatother

\usepackage{times,psfrag}
\usepackage{amsmath}
\usepackage{amsfonts}
\usepackage{amssymb}
\usepackage{hyperref}
\newcommand{\bq}{\begin{equation}}
\newcommand{\eq}{\end{equation}}
\newcommand{\bea}{\begin{eqnarray}}
\newcommand{\eea}{\end{eqnarray}}

\newcommand{\dd}{\mathrm{d}}
\newcommand{\ee}{\mathrm{e}}

\newcommand{\bbR}{\mathbb{R}}

\DeclareMathOperator{\SU}{\mathit{SU}}

\DeclareMathOperator{\SL}{\mathit{SL}}

\DeclareMathOperator{\spin}{\mathit{spin}}

\DeclareMathOperator{\tr}{tr}

\begin{document}

\title{Higher curvature Bianchi identities, generalised geometry and $L_{\infty}$ algebras}
\author{Andr\'e Coimbra}
\email{andre.coimbra@aei.mpg.de}
\affiliation{Max Planck Institute for Gravitational Physics (Albert Einstein Institute)\\
Am M\"uhlenberg 1, D-14476 Potsdam, Germany}

\begin{abstract}
The Bianchi identities for bosonic fluxes in supergravity can receive higher derivative quantum and string corrections, the most well known being that of Heterotic theory $\dd H = \tfrac14\alpha'(\tr F^2 - \tr R^2)$. Less studied are the modifications at order $R^4$ that may arise, for example, in the Bianchi identity for the seven-form flux of M theory compactifications. We argue that such corrections appear to be incompatible with the exceptional generalised geometry description of the lower order supergravity, and seem to imply a gauge algebra for the bosonic potentials that cannot be written in terms of an (exceptional) Courant bracket. However, we show that this algebra retains the form of an $L_{\infty}$ gauge field theory, which terminates at a level ten multibracket for the case involving just the seven-form flux.
\end{abstract}

\maketitle

\section{Generalised geometry and Bianchi identities}
The formalism of generalised geometry has proven to be a very powerful tool to tackle problems in string theory and supergravity. By looking at structures on a generalised tangent space which has `baked in' the much richer gauge field content of these theories, it provides a unified language for the bosonic sector that brings within reach previously intractable problems. However, precisely because the gauge fields are built into the definition of the generalised geometry, their Bianchi identities are assumed  by construction and any modification of them requires a change in the formalism. As we will review shortly, including the first $\alpha'$ correction due to the Green-Schwarz mechanism in Heterotic requires relaxing the exactness condition of the Courant algebroid in `base' generalised geometry, and adding the Ramond-Ramond fields (or moving to M theory) requires the introduction of an exceptional Courant algebroid. In this paper we will argue that further considering $R^4$ corrections -- which would be highly desirable as it could provide a path to finally obtain their supersymmetric completion and would have applications to phenomenological models that rely on perturbative effects to fix moduli in flux compactifications  -- implies again expanding the exceptional generalised geometry, and also that the gauge algebra can no longer be captured by a bracket acting just on  elements of the generalised tangent space. Finally we will show that there is nonetheless an $L_{\infty}$ algebra structure remaining, which we compute explicitly for a particular case.

\subsection{Generalised geometry}

Generalised geometry was originally introduced in~\cite{Hitchin:2004ut, gualtieri} as a way of combining complex and symplectic geometry, by considering structures on the generalised tangent bundle $E$
\begin{equation}\label{eq:exact-Courant}
T^* \rightarrow E \rightarrow T,
\end{equation}
so that $E$ is locally isomorphic to the sum of the tangent and cotangent bundle, $E\sim T\oplus T^*$. The generalised tangent bundle is naturally equipped with a Dorfman bracket or, equivalently, its antisymmetrisation a Courant bracket~\cite{dorfman, courant}
\begin{equation}\label{eq:Courant}
 [X_1,X_2]
    = [x_1,x_2] + {\cal L}_{x_1} \lambda_2 - {\cal L}_{x_2} \lambda_1
       - \tfrac12 \dd  ( i_{x_1} \lambda_2 -i_{x_2} \lambda_1 ) ,
\end{equation}
where $X_i= x_i+\lambda_i \in E$. The generalised tangent bundle is in fact an example of an exact Courant algebroid~\cite{Liu:1995lsa} and it possesses a three-form $H$ that is closed~\cite{severa}, 
\begin{equation}
\dd H = 0,
\end{equation}
which can be thought of as the curvature of a ``gerbe'' $B$~\cite{Hitchin:2005in}, i.e. $H= \dd B$ locally, which specifies a splitting of the sequence~\eqref{eq:exact-Courant}. Physicists quickly realised that this formalism provides a way of geometrising the NSNS sector of type II supergravity~\cite{Lindstrom:2004eh, Lindstrom:2004iw, Grana:2004bg, Grana:2005sn}, $B$ being  identified with the Kalb-Ramond field and $H$ being its flux. The Dorfman bracket along a generalised vector $L_X, X\in E$ then generates the combined (infinitesimal) bosonic symmetries of the theory: diffeomorphisms $\mathcal{L}_x$ by taking the Lie derivative along a vector field $x\in T$, and gauge $B$-shifts by $\dd\lambda$, exact two-forms parametrised by one-forms $\lambda\in T^*$. Introducing also a metric, it is possible to unify all the NSNS fields into a single object, and rewrite all the supergravity equations as a generalised geometry equivalent of Einstein gravity~\cite{Coimbra:2011nw} (see also~\cite{Aldazabal:2013sca} for an overview of the closely related subject of Double Field Theory that often implies many of these results).

\subsection{Heterotic generalised geometry}

In Heterotic theory, however, the field strength $H$ is no longer closed. Supersymmetry and the Green-Schwarz anomaly cancelation mechanism~\cite{Green:1984sg} require that $H$ satisfy a more complicated Bianchi identity.  This can be handled in the generalised geometry formalism by enlarging the generalised tangent space.  The resulting ``Heterotic generalised geometry''~\cite{Garcia-Fernandez:2013gja, Baraglia:2013wua, Anderson:2014xha, delaOssa:2014cia, het-gg, Bedoya:2014pma} is given in terms of a bundle which is a transitive, but not exact, Courant algebroid $E$, that can be built as a result of two extensions
\begin{equation}
\begin{aligned}
\mathfrak{g} \rightarrow \mathcal{A} \rightarrow T, \\
T^* \rightarrow E \rightarrow \mathcal{A} .
\end{aligned}
\end{equation} 
The first sequence defines a Lie algebroid $ \mathcal{A}$ known as the Atiyah algebroid for the quadratic Lie algebra  $\mathfrak{g}$, which replaces the role of the tangent bundle $T$ in the original generalised geometry. Writing $X_i= x_i+\Lambda_i+\lambda_i \in E \sim T\oplus\mathfrak{g}\oplus T^*$, the Courant bracket takes the form
\begin{equation}\label{eq:HetCourant}
\begin{split}
 [X_1,X_2]
    = & [x_1,x_2]  +  {\cal L}_{x_1} \Lambda_2 - {\cal L}_{x_2} \Lambda_1 + [\Lambda_1,\Lambda_2] \\&+ {\cal L}_{x_1} \lambda_2 - {\cal L}_{x_2} \lambda_1
       - \tfrac12 \dd  ( i_{x_1} \lambda_2 -i_{x_2} \lambda_1 ) +  \tr (\Lambda_2 \dd \Lambda_1 - \Lambda_1 \dd \Lambda_2) .
\end{split}
\end{equation}
The bundle $E$ then encodes the information for local gauge fields: a two-form\footnote{The heterotic $B$ field we are considering here is not gauge invariant under Yang-Mills transformations and is not a gerbe connection, it is a rather more complicated object~\cite{Witten:1999eg}.} $B$ and a Yang-Mills one-form $A$ taking values in $\mathfrak{g}$. These fields are not independent, they satisfy the global condition in terms of their respective field-strengths 
\begin{equation}
\dd H = \tr F^2.
\end{equation} 
By considering a Lie group $G$ (with algebra $\mathfrak{g}$) composed of two factors, a ``gravitational'' Lorentz group, and the usual $SO(32)$ or $E_8\times E_8$ (and choosing the correct normalisation of the metric in $\mathfrak{g}$), one obtains the Heterotic Bianchi identity:
\begin{equation}
\label{eq:het-bianchi}
\dd H = \tfrac14\alpha'(\tr F^2 - \tr R^2),
\end{equation}
where $R$, the field strength of the Lorentz factor, is now identified with the gravitational curvature. Once more, the Courant bracket in $E$ precisely reproduces the physical infinitesimal  bosonic symmetries: diffeomorphims  $\mathcal{L}_x$, $B$-shifts by  $\dd\lambda$, and  now also non-Abelian gauge transformations by some parameter $\Lambda\in\mathfrak{g}$. It is then possible to show that formulating the generalised equivalent of Einstein gravity in $E$ precisely reproduces the known Heterotic supergravity to order $\alpha'$~\cite{Garcia-Fernandez:2013gja}. The `trick' of treating the gravitational term in~\eqref{eq:het-bianchi} as if it were a Yang-Mills factor goes back to~\cite{Bergshoeff:1988nn}, though, as shown there, supersymmetry requires that the $\tr R^2$ be given by the curvature of a specific torsionful connection $\nabla - \tfrac12 H$. In~\cite{het-gg} it was shown that this is entirely consistent with the generalised geometry set-up.

\subsection{M theory and $E_{7(7)}\times\bbR^+$ generalised geometry}
\label{sec:e7}

In M theory, the equation of motion for the four-form flux $\mathcal{F}$ in eleven-dimensional supergravity~\cite{Cremmer:1978km} is corrected by higher order terms, starting with eight derivatives~\cite{Vafa:1995fj, Duff:1995wd}
\begin{equation}
\label{eq:11d-Feom}
\dd * \mathcal{F} =-\tfrac12 \mathcal{F}^2 + \kappa(\tr R^4 - \tfrac14 ( \tr R^2 )^2 ),
\end{equation}
where $\kappa$ is some constant which will be set to 1 as it will not influence the rest of our discussion, and with further terms which are functions of the flux expected to appear at the same order in derivatives but whose complete form is not yet known.

In order to find four-dimensional Minkowski backgrounds of M theory, one considers field ans\"atze that are compatible with the external global Lorentz symmetry. This means decomposing the eleven-dimensional manifold $\mathcal{M}_{11}$ as a warped product 
\begin{equation}
 \dd s^2_{11}(\mathcal{M}_{11}) = \ee^{\Delta}\dd s_4^2(\bbR^{3,1})+ \dd s^2_7 (M),
\end{equation} 
 where $M$ is some seven-dimensional internal space and $\Delta$ the warp factor, demanding that the fields depend only on internal coordinates, and keeping the components of the  $\mathcal{F}$ flux which are external scalars, i.e. the purely internal four-form $F$ and seven-form $\tilde{F}$. Their components are set in terms of the eleven-dimensional $\mathcal{F}$ simply by restricting
\begin{equation}
F = \mathcal{F}|_{M}  ,\quad  \tilde{F} = (*\mathcal{F})|_{M},
\end{equation} 
where $*\mathcal{F}$ is the eleven-dimensional Hodge dual. All other components of $\mathcal{F}$ are set to zero. The fact that $\mathcal{F}$ is closed in eleven dimensions together with the equation of motion~\eqref{eq:11d-Feom} then imply the Bianchi identities for the internal fluxes
\begin{align}
&\dd F = 0, \\[5pt] \label{eq:tF-bianchi}
&\dd \tilde{F} = -\tfrac12 F^2 + \tr R^4 - \tfrac14 ( \tr R^2 )^2 ,  
\end{align}
where the second equation should be taken as purely formal, since it vanishes identically in the seven-dimensional $M$. These induce internal local potentials, a three-form $C$ and a six-form $\tilde{C}$, which together with a Riemannian metric for $M$ and a warp factor $\Delta$ make up the bosonic degrees of freedom of the theory. 

Ignoring the higher-curvature terms, it was shown in~\cite{Coimbra:2012af} that this supergravity set-up (together with the fermionic sector) has a very natural interpretation as the analogue of Einstein gravity when formulated in $E_{7(7)}\times\bbR^+$ generalised geometry, also known as exceptional generalised geometry~\cite{Hull:2007zu,Pacheco:2008ps,e7gg}.  One introduces a generalised tangent bundle again as a series of extensions, such that it has a local form
\begin{equation}\label{eq:e7-gtb}
E \sim T\oplus \Lambda^2T^* \oplus \Lambda^5 T^* \oplus \left( T^*\otimes \Lambda^7T^* \right) ,
\end{equation}
which encodes the bosonic symmetries of the theory, namely diffeomorphism generated by vector fields $x$ and shifts by two-forms $\omega$ and five-forms $\sigma$ of the gauge fields $C$ and $\tilde{C}$ respectively. The peculiar one-form-tensor-seven-form term $\tau\in T^*\otimes \Lambda^7T^* $ would be a charge for a ``dual graviton'' field which happens to vanish identically in seven-dimensional  compactifications and so, while it is implied by the higher-dimensional M theoretic geometry, it has no immediate physical meaning given this setup (though see~\cite{Hohm:2014fxa}, for example, where $\tau$ and other mixed-symmetry charges become crucial to formulate higher exceptional geometries). By construction, the generalised tangent bundle defines a global closed four-form $F$ that can locally be expressed in terms of the potential 
\begin{equation}
F= \dd C,
\end{equation} 
and a seven-form such that  
\begin{equation}
\tilde{F}= \dd \tilde{C} -\tfrac12 CF .
\end{equation}
The supergravity Bianchi identities inherited from eleven-dimensions are thus automatically satisfied. The gauge algebra is then given by the natural differential structure over $E$, the (exceptional) Courant bracket of two generalised vector fields, which takes the form\footnote{The $j$-notation corresponds to a projection to the $T^*\otimes \Lambda^7T^*$ space, see~\cite{Pacheco:2008ps,e7gg} for its precise definition, though it will not be needed for what follows.}
\begin{equation}
\begin{split}
[X_1,X_2]& = \mathcal{L}_{x_1}x_2 
       +  \mathcal{L}_{x_1} \omega_2  - \mathcal{L}_{x_2} \omega_1 -\tfrac12 \dd (i_{x_1} \omega_2  - i_{x_2} \omega_1) \\ 
       &+   \mathcal{L}_{x_1} \sigma_2 - \mathcal{L}_{x_2} \sigma_1 -\tfrac12 \dd (i_{x_1} \sigma_2  - i_{x_2} \sigma_1)+\tfrac12 \sigma_1\dd\sigma_2 -\tfrac12 \sigma_2\dd\sigma_1
      \\& +   \tfrac12\mathcal{L}_{x_1} \tau_2- \tfrac12\mathcal{L}_{x_2} \tau_1
          +\tfrac12 ( j\sigma_1\wedge\dd\omega_2 -  j\sigma_2 \wedge\dd\omega_1 )
          - \tfrac12 ( j\sigma_2\wedge\dd\omega_1 - j\sigma_1 \wedge\dd\omega_2 ),
\end{split}
\end{equation}
where we write $X_i = x_i+\omega_i+\sigma_i+\tau_i \in E $. The bundle $E$ has a natural $E_{7(7)}\times\bbR^+$ structure and the bracket is compatible with this structure. The bosonic degrees of freedom turn out to simply be the components of a generalised metric for the generalised tangent space, reducing the structure group to its maximal compact subgroup $\SU(8)/\mathbb{Z}_2$, and the corresponding generalised Ricci scalar precisely reproduces the supergravity bosonic action. That eleven-dimensional supergravity admitted this larger symmetry had already been proven in~\cite{deWit:1986mz}. This efficient rewriting  has made it possible to tackle several physical problems in full generality (without needing to restrict to some subsector of the fluxes, for example), such as classifying supersymmetric backgrounds~\cite{Grana:2011nb,Grana:2012ea,Coimbra:2014uxa, Ashmore:2015joa, Coimbra:2015nha, Ashmore:2016qvs,Coimbra:2016ydd, Coimbra:2017fqv, Malek:2017njj} or describing their moduli spaces and holographic duals~\cite{Ashmore:2016oug, Ashmore:2018npi}. It would thus seem promising to apply the same techniques with the higher derivative corrections included~\cite{Coimbra:2017fqj}.

\subsection{M theory corrections}
\label{sec:7-corr}

So now let us consider adding back the higher curvature terms originating in eleven dimensions~\eqref{eq:11d-Feom}. These are incompatible with the $E_{7(7)}\times\bbR^+$ generalised tangent bundle previously introduced, since by construction it forces  $\tilde{F}=  \dd \tilde{C} -\tfrac12 CF$. On the contrary, the corrected Bianchi identity~\eqref{eq:tF-bianchi} implies the local form for the flux $\tilde{F}$
\begin{equation}\label{eq:tF}
 \tilde{F} = \dd \tilde{C} -\tfrac12 C   F  +\omega_7(A) -\tfrac14 \omega_3(A)   \tr R^2 ,
\end{equation}
where $A$ is the spin-connection for the Riemann curvature $R$ and $\omega_{n}(A)$ denotes the Chern-Simons $n$-form for $A$ such that $\dd\omega_{2n-1}(A) = \tr R^{n} $, see the appendix for their explicit form.

Following the same trick as for the Heterotic case, we may treat at first the curvature $R$ simply as the field strength for a generic Yang-Mills gauge field $A$ taking values in some algebra $\mathfrak{g}$, though naturally it will eventually be necessary to identify $\mathfrak{g}$ with $\spin(7)$ and express $A$ in terms of gravitational degrees of freedom.\footnote{Though an intriguing possibility is to consider a larger gauge group that could accommodate the flux degrees of freedom, such as taking $\mathfrak{g} = su(8)$ and relating $A$ to the $SU(8)$ connections implied by supersymmetry. This could naturally give rise to a Bianchi identity which includes higher derivative flux terms.} The Heterotic generalised geometric prescription would then lead us to consider structures over a generalised tangent space of the form
\begin{equation}
T\oplus \mathfrak{g} \oplus \Lambda^2T^* \oplus \Lambda^5 T^* \oplus T^*\otimes \Lambda^7T^* ,
\end{equation}
in other words, replacing the tangent bundle component of~\eqref{eq:e7-gtb} with the Atiyah algebroid.

In what follows, however, we will restrict ourselves to simpler versions of this problem, which will still suffice to show that the situation is more complex than the one of Heterotic generalised geometry. In particular, we will find gauge algebras that are best described in terms of higher order $L_{\infty}$-algebras.

In section~\ref{sec:r3} we will first look at a Bianchi identity
\begin{equation}
\dd \tilde{F}_5 = \tr R^3,
\end{equation} 
where $\tilde{F}_5$ is a five-form which, even though it has no immediate physical motivation, is easier to handle and already displays the important features we wish to demonstrate. The corresponding generalised tangent space will be of the form
\begin{equation}
T\oplus \mathfrak{g}\oplus \Lambda^3 T^*.
\end{equation}

We will then move on in section~\ref{sec:r4} to the case 
\begin{equation}
\dd \tilde{F}_7 = \tr R^4 ,
\end{equation} 
where now $\tilde{F}_7$ is genuinely a seven-form, and so this corresponds to a special case of~\eqref{eq:tF}. The generalised tangent space is then 
\begin{equation}
T\oplus \mathfrak{g}\oplus \Lambda^5 T^*.
\end{equation}

In both cases we will find that the Bianchi identities imply a gauge algebra which cannot be expressed in terms of simply a Courant bracket. Instead it is of the type of the $L_{\infty}$ field theory formalism of~\cite{hohm-zwiebach}. The analysis of the complete Bianchi identity implied by the corrected eleven-dimensional supergravity will be left for future work.

As an aside, we expect that similar conclusions would hold for $(2n-1)$-form fluxes $\tilde{F}_{(2n-1)}$ satisfying
\begin{equation}\label{eq:bianchi-rn}
\dd \tilde{F}_{(2n-1)} = \tr R^n,
\end{equation} 
based on generalised tangent spaces
\begin{equation}
E\sim T\oplus \mathfrak{g}\oplus \Lambda^{_{(2n-3)}}T^*,
\end{equation}
though we will not attempt to prove this here. Note as well that in all these cases the Bianchi identities, when viewed in cohomology classes, correspond to obstructions to this construction, namely the requirement that $n$-th Chern character of the gauge vector bundle is trivial.

We also remark that the fact that the gauge algebras we are examining fit into the $L_{\infty}$ setting is not surprising. It has already been shown that the ``higher Courant algebroids'' of the type $T\oplus\Lambda^pT^*$ have an associated  $L_{\infty}$ algebra~\cite{zambon}, and the extra terms we are considering arise from adding an invariant polynomial to the Bianchi, which in the context of chiral anomalies lead to the well-known ``descent equations'' derived from the extended Cartan homotopy~\cite{Manes:1985df}, with many of the terms in the brackets we present here being directly related to the extra homotopy operator.

\subsection{$L_{\infty}$ algebras and field theory}
\label{sec:linf}

$L_{\infty}$ algebras or strong homotopy Lie algebras, introduced in~\cite{Zwiebach:1992ie, Lada:1992wc} to the physics context, have found numerous applications in both mathematics and physics, see~\cite{Stasheff:2018vnl} for a recent review of the field. In particular, they can be found in the theories of Courant algebroids and generalised geometry. Courant algebroids were shown to have an $L_3$-algebra in~\cite{roytenberg-weinstein}. In the case of Heterotic Courant algebroids, it has recently been proven that this algebra is directly connected to the physical problem of finding the moduli of finite deformations of the Strominger-Hull system~\cite{Ashmore:2018ybe}. Higher Courant algebroids over a space $T\oplus\Lambda^pT^*$ were proven to have $L_{\infty}$ algebras for arbitrary $p$ in~\cite{zambon} using a derived bracket construction, and in~\cite{Baraglia:2011dg} a large class of ``Leibniz algebroids'' (a Leibniz bracket being a generalisation of the Lie derivative that still satisfies the Leibniz identity but is not necessarily anti-symmetric), of which exceptional generalised geometries are examples, were likewise shown to admit $L_{\infty}$ algebras. This later point was further explored in~\cite{Arvanitakis:2018cyo,Arvanitakis:2019cxy}, where the correct $L_{\infty}$ algebra was demonstrated to follow from interpreting the M theory geometries as dg-symplectic manifolds. More broadly speaking, the higher structures that feature in string and M theory are known to be classified by super homotopy theory, see the review~\cite{Fiorenza:2019ckz} and references therein. In particular, note that the Heterotic generalised geometry that we described earlier corresponds to a ``string Lie
2-algebra''~\cite{Sati:2009ic}, while anomaly cancelation in M theory was examined in this formalism in~\cite{Fiorenza:2019usl}.  There has also been much current work showing how such structures  appear in the related fields of Double/Exceptional Field Theory, for example in~\cite{Deser:2014mxa, Wang:2015hca, Deser:2016qkw, Lavau:2017tvi, Hohm:2017cey, Deser:2017fko, Cederwall:2018aab, Hohm:2018ybo, Cagnacci:2018buk, Deser:2018flj, Hohm:2019wql, Bonezzi:2019ygf, Lavau:2019oja}.

Recently, in~\cite{hohm-zwiebach} (see also~\cite{Jurco:2018sby}) many of these ideas were systematised in a manner to be more immediately applicable to physics, by introducing the notion of ``$L_{\infty}$ gauge field theories''. It is this approach that we will be following, and we start by quickly reviewing some of the concepts that will be relevant here.

There are a few alternative ways of defining an $L_{\infty}$ algebra. Following the conventions of~\cite{hohm-zwiebach} we will be working with the ``$\ell$-picture'' in terms of graded-antisymmetric multilinear brackets. Given a $\mathbb{Z}$-graded vector space
\begin{equation}
V= \oplus_{i\in\mathbb{Z}}V_i ,
\end{equation}
where the subscript denotes the degree, one defines an $L_{\infty}$ algebra by endowing it with a series of multilinear products $\ell_n :\, \Lambda^n V \mapsto V $. These brackets are of degree $n-2$,  i.e. for inputs $v_i \in V$, the total degree of $\ell_n(v_1,\dots , v_n)$ is
\begin{equation}\label{eq:ln-degree}
\text{deg}\,\ell_n(v_1,\dots , v_n) = n -2 +\sum_{i=1}^n \deg v_i .
\end{equation}
They are also graded antisymmetric,
\begin{equation}\label{eq:Linf-antisym}
\ell_n(v_{\sigma(1)},\dots,v_{\sigma(n)}) = (-1)^{|\sigma |}\epsilon(\sigma)\ell_n(v_1,\dots,v_n) ,
\end{equation}
for some permutation $\sigma$ and where $\epsilon$ is the Koszul sign for the given permutation and grading of $V$. Crucially, for each $n$ the brackets must also satisfy a Jacobi identity ``up to higher homotopies'', namely the generalised Jacobi identities 
\begin{equation}\label{eq:gen-Jac}
\sum_{i+j= n+1}(-1)^{i(j-1)}\sum_{\sigma}(-1)^{|\sigma |}\epsilon(\sigma)\ell_j(\ell_i(v_{\sigma(1)},\dots,v_{\sigma(i)}),v_{\sigma(i+1)},\dots,v_{\sigma(n)}) = 0,
\end{equation}
or in abbreviated form
\begin{equation}\label{eq:simple-gen-Jac}
\sum_{i+j= n+1}(-1)^{(j-1)i}\ell_j\ell_i = 0 .
\end{equation}
Explicitly, this gives at level one
\begin{equation}
\ell_1(\ell_1(v)) = 0,
\end{equation}
which shows that the graded vector space $V$ of an $L_{\infty}$ algebra forms a chain complex with the operator $\ell_1$. Level two establishes $\ell_1$ as a derivation on $\ell_2$,
\begin{equation}
\ell_1(\ell_2(v_1,v_2)) = \ell_2(\ell_1(v_1),v_2) +(-1)^{|v_1|} \ell_2(v_1,\ell_1(v_2)).
\end{equation}
Level three would be the `traditional' Jacobi identity, where the $L_{\infty}$ algebra starts to diverge from the normal graded Lie algebras
\begin{equation}
\begin{split}
&\ell_1(\ell_3(v_1,v_2,v_3)) + \ell_3(\ell_1(v_1),v_2,v_3)  + (-1)^{|v_1|}\ell_3(v_1,\ell_1(v_2),v_3)  \\&\quad + (-1)^{|v_1|+|v_2|}\ell_3(v_1,v_2,\ell_1(v_3)) + \ell_2(\ell_2(v_1,v_2),v_3)\\&\quad +(-1)^{(|v_2|+|v_3|)|v_1|}\ell_2(\ell_2(v_2,v_3),v_1) +(-1)^{(|v_1|+|v_2|)|v_3|}\ell_2(\ell_2(v_3,v_1),v_2)  = 0 ,
\end{split}
\end{equation}
and so on.

Proceding with the proposal of~\cite{hohm-zwiebach} for a gauge field theory, one considers spaces of type\footnote{In~\cite{hohm-zwiebach} an extra subspace $V_{-2}$ is also allowed,  corresponding to the equations of motion, but we will not make use of it here.}
\begin{equation}\label{eq:Linf-vector}
V= \oplus_{i>0}V_i\oplus V_0\oplus V_{-1} .
\end{equation}
An important point here is that, since one allows a space with negative grading, there is a priori no guarantee that the $L_{\infty}$ algebra will ever terminate even for a finite number of $V_i$. This is in contrast to $L_n$-algebras, defined such that the graded vector space is concentrated in degrees 0 to $n-1$ and therefore all brackets of degree higher than $n+1$ vanish trivially as a consequence of~\eqref{eq:ln-degree}. However, we will see that the cases we consider in the next sections do indeed truncate and there is a finite number of brackets to consider.

In order to find the physical meaning of~\eqref{eq:Linf-vector}, one identifies elements $X\in V_0$ with gauge parameters and $\Psi\in V_{-1}$ are taken to be the gauge fields. Elements of $\oplus_{i>0}V_i$ are to be thought of as making up a tower of trivial gauge parameters. An $L_{\infty}$ gauge field theory may then be defined with the symmetries given by
\begin{equation}\label{eq:Linf-symm}
\delta_X \Psi = \Sigma_n \frac{1}{n!} (-1)^{\tfrac{n(n-1)}{2}} \ell_{n+1}(X,\Psi^n) ,
\end{equation}
satisfying a gauge algebra
\begin{equation}\label{eq:Linf-gauge}
[\delta_{X_1},\delta_{X_2}] \Psi = \delta_{[X_1,X_2]}\Psi,\quad [X_1,X_2] = \Sigma_n \frac{1}{n!} (-1)^{\tfrac{n(n-1)}{2}} \ell_{n+2}(X_1,X_2,\Psi^n) .
\end{equation}
Note in particular that in this formalism the gauge algebra of the parameters is permitted to depend explicitly on the fields. In what follows we will show how the higher curvature problem we are considering fits precisely into this picture.

We also remark that another way the gauge fields may appear in Courant brackets is via their `twisting'. One can think of the fields as defining connections that split the exact sequences that define the generalised tangent bundle, i.e. they make explicit the isomorphism $E\sim T\oplus \dots$. Under that map, the Courant bracket then becomes twisted by the curvature of the connection -- for example, for the exact Courant algebroid case the $B$ field splits the defining sequence $T^*\rightarrow E \rightarrow T$ through a map $\ee^B: x + \lambda \mapsto x + \lambda + i_x B$ (see~\cite{gualtieri}), and the bracket~\eqref{eq:Courant} then gets modified by $[\ee^B X_1, \ee^B X_2] = \ee^B[X_1, X_2] + i_{x_1}i_{x_2} H$. Note also that if $B$ is closed, physically `pure gauge', then it is a symmetry of the bracket, extending the usual diffeomorphism invariance of the Lie bracket. One would expect that something similar for the higher order brackets we define in the following sections, it should be possible to twist them with the field strengths of the gauge fields, and pure gauge finite transformations should leave them invariant. However, we do not attempt to verify this here, a full study of the twisted structure of the higher order generalised tangent bundles, their patching rules (which would involve computing finite gauge transformations of the Chern-Simons forms), the automorphisms of these $L_{\infty}$ structures, etc. will be left for future work.

\section{$\dd \tilde{F}_5 = \tr R^3$}
\label{sec:r3}

We begin by considering a theory with a globally defined five-form flux $\tilde{F}$ and a Yang-Mills  $\mathfrak{g}$-valued potential $A$ with corresponding field strength $R$ such that
\begin{align}
&\dd_A R = 0 ,\\[5pt]
&\dd \tilde{F}_5 = \tr R^3 .
\end{align}
We can thus define a four-form potential $\tilde{C}$ for the flux by 
\begin{equation}
\tilde{F}_5 = \dd \tilde{C}_4 + \omega_5(A).
\end{equation} 
Much like the $B$ field in Heterotic theory, we find that since $\tilde{F}_5$ is gauge invariant,  $\tilde{C}_4$ must transform to compensate for a variation of the Chern-Simons five-form $\omega_5(A)$. That is, if $\Lambda \in \mathfrak{g}$ parametrises an infinitesimal gauge transformation, we must have that locally
\begin{equation}
\dd \delta_{\Lambda} \tilde{C}_4 = -\delta_{\Lambda} \omega_5(A) = -\dd \omega^1_4(\Lambda,A) = -  \dd \tr  \dd \Lambda \left(  A \dd A  + \tfrac12 A^3   \right) ,
\end{equation}
from the properties of the Chern-Simons forms (see appendix). It is also clear that $\tilde{F}$ remains invariant under  shifts of $\tilde{C}_4$ by a closed four-form, locally parametrised by the exterior derivative of some three-form $\sigma$. Together with a diffeomorphism symmetry parametrised by some vector field $x$, we have that the potentials obey the infinitesimal gauge transformations
\begin{align}\label{eq:5-symm}
\begin{split}
\delta_X A &= \mathcal{L}_x A - \dd \Lambda - [A,\Lambda],\\[5pt]
\delta_X \tilde{C}_4 &= \mathcal{L}_x \tilde{C}_4 - \dd \sigma -  \tr  \dd \Lambda \left(  A \dd A  + \tfrac12 A^3   \right) , 
\end{split}
\end{align}
where $\delta_X$ denotes the combined infinitesimal diffs, gauge and shifts in terms of parameters $ X = x + \Lambda + \sigma$. This therefore suggests a generalised tangent space
\begin{equation}
E = T \oplus \mathfrak{g} \oplus \Lambda^3 T^* .
\end{equation}

So far, this precisely matches the procedure for constructing the Heterotic generalised geometry, see for example~\cite{het-gg}. However, let us look at how the algebra of transformations $\delta_X$ closes when acting on the fields. Taking two parameters $X_1,X_2\in E$, we find that
\begin{align}
\begin{split}
[ \delta_{X_1} ,\delta_{X_2} ] A &= \mathcal{L}_{[x_1, x_2]} A -\dd \left( [\Lambda_1,\Lambda_2] + i_{x_1} \dd \Lambda_2 -  i_{x_2} \dd \Lambda_1 \right)\\  
& -[A, [\Lambda_1,\Lambda_2] + i_{x_1} \dd \Lambda_2 -  i_{x_2} \dd \Lambda_1 ],\\[5pt]
[\delta_{X_1},\delta_{X_2}] \tilde{C}_4&=   \mathcal{L}_{[x_1, x_2]} \tilde{C}_4 - \dd (i_{x_1} \dd \sigma_2 - i_{x_2} \dd \sigma_1 +\tfrac12 \dd i_{x_1} \sigma_2 -\tfrac12 \dd i_{x_2} \sigma_1) \\  
&   -  \tr\dd \left( [\Lambda_1,\Lambda_2] + i_{x_1} \dd \Lambda_2 -  i_{x_2} \dd \Lambda_1 \right)  \left( A (\dd A )^2 +\tfrac12 A^3 \right)\\
&+\dd \tr  \left( \Lambda_1 \dd \Lambda_2 \dd A -   \Lambda_2 \dd \Lambda_1 \dd A \right),
\end{split}
\end{align}
so we have that indeed the algebra closes on a parameter $X_3$ given by
\begin{align} \label{eq:5-gauge}
\begin{split}
[\delta_{X_1},\delta_{X_2}](A+ \tilde{C}_4) &= \delta_{X_3} (A + \tilde{C}_4),\\[5pt]
X_3 &= [x_1, x_2] +  [\Lambda_1,\Lambda_2] + i_{x_1} \dd \Lambda_2 -  i_{x_2} \dd \Lambda_1  \\
& +i_{x_1} \dd \sigma_2 - i_{x_2} \dd \sigma_1 +\tfrac12 \dd i_{x_1} \sigma_2 -\tfrac12 \dd i_{x_2} \sigma_1 \\
&-\tr\big( \Lambda_1 \dd \Lambda_2 \dd A -   \Lambda_2 \dd \Lambda_1 \dd A \big)\in E,
\end{split}
\end{align}
but note that this depends not just on $X_1$ and $X_2$ but also explicitly on the fields. Therefore, unlike the previous examples in generalised geometry, the gauge algebra does not define for us a bracket over just the space $E$. It does, nonetheless, fit into the $L_{\infty}$ field theory setting.

\subsection{An $L_{\infty}$ gauge algebra for $R^3$}

Let us then introduce the graded vector space:
\begin{equation}
V = V_3\oplus V_2\oplus V_1 \oplus V_0 \oplus V_{-1} ,
\end{equation}
where\footnote{A more `generalised' treatment in the sense of~\cite{e7gg} would presumably involve introducing a space of ``generalised frames'' for $E$ (that is a subspace of $\text{End}(E)$ that preserves the defining generalised structures -- $O(d,d)$ in NSNS generalised geometry, $E_{7(7)} $ in exceptional generalised geometry, etc.), and identifying its `geometric subspace'  with $V_{-1}$ which is used to construct the physical brackets.}
\begin{align}
\begin{split}
 &V_3 = C^{\infty}(M),\quad V_2 =   T^*  ,\quad V_1= \Lambda^2 T^* ,\\[5pt]
  &V_0 = E = T\oplus \mathfrak{g} \oplus \Lambda^3 T^* ,\quad   V_{-1}= T^* \otimes\mathfrak{g} \oplus \Lambda^4 T^* ,
\end{split}
\end{align}
and we label elements in the subspaces as
\begin{equation}
\xi \in V_3\oplus V_2\oplus V_1, \quad X= x + \Lambda + \sigma \in V_0,\quad  \Psi = A + \tilde{C} \in V_{-1} .
\end{equation}

We will then endow $V$ with a series of multilinear brackets to define an $L_{\infty}$ algebra that will realise the gauge algebra~\eqref{eq:5-gauge}. Terms in the brackets involving only elements in $V_{i>0}$  or the vector + three-form part of $V_0$  will be necessarily the ones in~\cite{zambon}, but we must introduce new products for terms involving the Lie algebra $\mathfrak{g}$.  Comparing with~\eqref{eq:Linf-gauge}, we can directly read off some of the multibrackets, since we must insist   that picking particular elements $A+\tilde{C}_4=\Psi\in V_{-1}$ corresponds to specifying the  data for the supergravity gauge fields, i.e. that they satisfy a gauge algebra
\begin{equation}
\delta_X \Psi  = \ell_1(X) + \ell_2(X, \Psi) -\tfrac12 \ell_3(X, \Psi, \Psi) -\tfrac16 \ell_4(X,\Psi,\Psi,\Psi),
\end{equation}
with
\begin{equation}
[\delta_{X_1},\delta_{X_2}] \Psi = \delta_{X_3} \Psi, \quad X_3 = \ell_2(X_1, X_2) + \ell_3(X_1, X_2, \Psi),
\end{equation}
such that it precisely matches~\eqref{eq:5-symm} and~\eqref{eq:5-gauge} respectively.

Several more brackets are necessary to complete the algebra, which can be obtained from the requirement that they satisfy the generalised Jacobi identities~\eqref{eq:gen-Jac}. It is possible to do this exhaustively term-by-term since, due to  both the grading of the vector space $V$ and the subdivisions inside $V_0$ and $V_{-1}$, many will vanish trivially. For example, we will see that all brackets $\ell_n$ of level $n>2$ whose image is in $V_{-1}$ actually only map to the four-form subspace, i.e. they are $\tilde{C}$-type objects. On the other hand,  the brackets of level $n>2$ that take an object in $V_{-1}$ as input are all independent of $\tilde{C}$. So chaining together those sets of brackets is trivial.

Note that due to the grading and symmetry properties of the $\ell_n$ brackets~\eqref{eq:Linf-antisym}, products involving multiple factors of $X_i$ will always have to be antisymmetrised, and products involving products of $\Psi_i$ will always have to be symmetrised. We denote this explicitly using the typical index notation of symmetrisers and antisymmetrisers.

We find that the (non-trivial) $L_{\infty}$ products are then:

at level one  
\begin{align}\label{eq:r3-l1}
\ell_1(\xi) = \dd \xi , \quad \ell_1(X) = -\dd \Lambda - \dd \sigma, \quad \ell_1(\Psi) = 0,
\end{align}

at level two 
\begin{subequations}\label{eq:r3-l2}
\begin{align}
\begin{split}
&\ell_2(X,\xi) = \tfrac12 \mathcal{L}_{x}  \xi ,
\end{split}\\[7pt]
\begin{split}
&\ell_2(X_1, X_2) = [x_1, x_2] + [\Lambda_1,\Lambda_2] + \mathcal{L}_{x_1} \Lambda_2 -   \mathcal{L}_{x_2}  \Lambda_1
\\&\quad +  \mathcal{L}_{x_1} \sigma_2 -  \mathcal{L}_{x_2} \sigma_1 -\tfrac12 \dd i_{x_1} \sigma_2 +\tfrac12 \dd i_{x_2} \sigma_1,
\end{split}\\[7pt]
\begin{split}
&\ell_2(X,\Psi) = \mathcal{L}_{x}\Psi  - [A , \Lambda],
\end{split}
\end{align}
\end{subequations}

at level three 
\begin{subequations}\label{eq:r3-l3}
\begin{align}
\begin{split}
&\ell_3( \xi, X_1, X_2) = -\tfrac16 ( i_{x_{[1}} \mathcal{L}_{x_{2]}} + i_{[x_1,x_2]} ) \xi,
\end{split}\\[7pt]
\begin{split}
&\ell_3(X_1, X_2, X_3) = -\tfrac12\big(i_{x_{[1}} \mathcal{L}_{x_2}+ i_{[x_{[1},x_2]}+ i_{x_{[1}} i_{x_2} \dd \big) \sigma_{3]} ,
\end{split}\\[7pt]
\begin{split}
&\ell_3(X_1, X_2, \Psi) = -2\tr \Lambda_{[1}\dd \Lambda_{2]} \dd A  ,
\end{split}\\[7pt]
\begin{split}
&\ell_3(X, \Psi_1, \Psi_2) =  2\tr \dd \Lambda A_{(1} \dd A_{2)}  ,
\end{split}
\end{align}
\end{subequations}

 at level four
\begin{subequations}\label{eq:r3-l4}
\begin{align}
\begin{split}
&\ell_4(X_1, X_2, X_3, X_4) = - 12 \tr \Lambda_{[1}\Lambda_2\Lambda_3\dd\Lambda_{4]} ,
\end{split}\\[7pt]
\begin{split}\label{eq:r3-l4-2}
&\ell_4(X_1, X_2, X_3, \Psi) =6 \tr \Lambda_{[1}\Lambda_2\dd\Lambda_{3]} A -\tfrac32 i_{x_{[1}}   \,\ell_3(X_2,X_{3]},\Psi) 
,
\end{split}\\[7pt]
\begin{split}
&\ell_4(X, \Psi_1, \Psi_2, \Psi_3) =  3 \tr \dd \Lambda A_{(1} A_2 A_{3)}  ,
\end{split}
\end{align}
\end{subequations}

 at level five
\begin{subequations}\label{eq:r3-l5}
\begin{align}
\begin{split}
&\ell_5(X_1, \dots,X_5) = -12 \tr\Lambda_{[1}\Lambda_2\Lambda_3\Lambda_4\Lambda_{5]}-  \tfrac{5}{2}    i_{x_{[1}}   \,\ell_4(X_2,X_3,X_4,X_{4]})  \\&\quad - \tfrac13 i_{x_{[1}}i_{x_2  \,}\ell_3(X_3,X_4,X_{5]}),
\end{split}\\[7pt]
\begin{split} 
&\ell_5(X_1, X_2, X_3, X_4,\Psi) =  
-2i_{x_{[1}}  \,\ell_4(X_2,X_3,X_{4]},\Psi) -2 i_{x_{[1}}i_{x_2}\ell_3(X_3,X_{4]},\Psi) ,
\end{split}
\end{align}
\end{subequations}

and finally at level six  
\begin{align}\label{eq:r3-l6}
\begin{split}
&\ell_6(X_1, \dots,X_5,\Psi) = \tfrac{5}{3}   i_{x_{[1}} i_{x_2}  \,\ell_4(X_3,X_4,X_{5]},\Psi)
 +\tfrac52  i_{x_{[1}} i_{x_2}  i_{x_{3}}   \,\ell_3(X_4,X_{5]},\Psi)  .
\end{split}
\end{align}

All other brackets vanish. Even though there is a large number of them, it should be clear from the form of  the non-trivial brackets that they are straightforward to obtain by iterating through the generalised Jacobi identities. As an example, consider the fourth level identity when the inputs are in  $(V_0)^3\otimes V_{-1}$. This will involve the bracket $\ell_4(X_1, X_2, X_3, \Psi)$ given in~\eqref{eq:r3-l4-2}, which cannot be read directly from the gauge algebra (and has not been previously derived in the literature), instead we infer it from the generalised Jacobi identity. We compute from the lower order brackets
\begin{equation}\begin{split}
& \ell_3(\ell_2(X_{[1},X_2),X_{3]},\Psi) =    -2\tr \mathcal{L}_{x_{[1}} ( \Lambda_2 \dd\Lambda_3)\dd A+2  \tr  \Lambda_{[1}\dd   \Lambda_2 \Lambda_{3]} \dd A     ,\\[3pt]
& \ell_3(X_{[1},X_2,\ell_2(X_{3]}, \Psi)) = -2\tr \Lambda_{[1}\dd \Lambda_2  \big( \mathcal{L}_{x_{3]}} \dd A - \dd[A,\Lambda_{3]}]\big)  ,\\[3pt]
& \ell_2(X_{[1},\ell_3(X_2,X_{3]},\Psi)) =  -2\mathcal{L}_{x_{[1}}  \tr\Lambda_2\dd\Lambda_{3]}\dd A    + \dd i_{x_{[1}} \tr\Lambda_2\dd\Lambda_{3]}\dd A    ,
\end{split}
\end{equation}
and noting that $\ell_4(\ell_1(X_1),X_2,X_3,\Psi)$ and  $\ell_4(X_1,X_2,X_3,\ell_1(\Psi))$ vanish identically, the generalised Jacobi identity then reads
\begin{equation}\begin{split}
& \ell_1(\ell_4(X_1,X_2,X_3,\Psi))   = 3\ell_4(\ell_1(X_{[1}),X_2,X_{3]},\Psi)  +3\ell_2(X_{[1},\ell_3(X_2,X_{3]},\Psi)) \\
& \quad-3\ell_3(X_{[1},X_2,\ell_2(X_{3]}, \Psi))-3\ell_3(\ell_2(X_{[1},X_2),X_{3]},\Psi)  \\&\quad=6 \dd \tr \Lambda_{[1}\Lambda_2\dd\Lambda_{3]} A +3  \dd i_{x_{[1}} \tr\Lambda_2\dd\Lambda_{3]}\dd A  
\\&\quad =6 \dd\tr \Lambda_{[1}\Lambda_2\dd\Lambda_{3]} A -\tfrac32 \dd i_{x_{[1}}   \,\ell_3(X_2,X_{3]},\Psi),
\end{split}
\end{equation}
which is consistent with~\eqref{eq:r3-l4-2}.

\section{$\dd \tilde{F}_7 = \tr R^4$}
\label{sec:r4}

The analysis for a seven-form flux follows in much the same way as the five-form case we just considered, it is simply more computationally intensive. We again introduce a $\mathfrak{g}$-valued one-form potential $A$ with field strength $R$, and a globally defined seven-form $\tilde{F}$ such that
\begin{equation}
\dd \tilde{F}_7 = \tr R^4 ,
\end{equation}
and so locally we define a six-form potential $\tilde{C}$ by
\begin{equation}
\tilde{F}_7 = \dd \tilde{C}_6 + \omega_7(A) .
\end{equation}
As previously remarked, this is a toy example for the supergravity theory of section~\ref{sec:e7} when one truncates equation~\eqref{eq:tF}. Now, gauge invariance of $\tilde{F}_7$ once again implies that $\tilde{C}_6$ must vary as
\begin{equation}
\begin{aligned}
\dd \delta_{\Lambda} \tilde{C}_6  &= -\delta_{\Lambda} \omega_7(A) = -\dd\omega^1_6(\Lambda ,	A)  \\&= -  \dd \tr  \dd \Lambda \big( A (\dd A )^2 + \tfrac25 ( A^3  \dd A +   \dd A A^3  + A^5) + \tfrac15 ( A^2 \dd A A  +   A \dd A A^2  ) \big),
\end{aligned}
\end{equation}
for some gauge parameter $\Lambda \in \mathfrak{g}$ (from equation~\eqref{eq:app-omega6} in the appendix). We also have the usual diffeomorphism $\mathcal{L}_x$ and shift symmetries $\dd \sigma$ generated by, respectively, a vector field $x\in T$ and a five-form $\sigma\in\Lambda^5 T^*$. We are thus led to consider a generalised tangent space
\begin{equation}
\begin{aligned}
E &= T\oplus \mathfrak{g} \oplus \Lambda^5 T^* ,\\[5pt]
X &= x + \Lambda + \sigma \in E .
\end{aligned}
\end{equation}
This is a close cousin of the $\SL(8,\bbR)\times\bbR^+$ ``half-exceptional'' generalised geometry  obtained by truncating the $E_{7(7)}\times\bbR^+$ case, as was described in~\cite{Strickland-Constable:2013xta}. We can then group the infinitesimal symmetries as
\begin{equation}\label{eq:7-symm}
\begin{aligned}
\delta_X &= \,\text{infinitesimal diffs, gauge and shifts}\\[5pt]
\delta_X A &= \mathcal{L}_x A - \dd \Lambda - [A,\Lambda]\\[5pt]
\delta_X \tilde{C}_6 &= \mathcal{L}_x \tilde{C}_6 - \dd \sigma \\ 
&-  \tr  \dd \Lambda \left( A (\dd A )^2 + \tfrac25 ( A^3  \dd A +   \dd A A^3  + A^5) + \tfrac15 ( A^2 \dd A A  +   A \dd A A^2  ) \right) 
\end{aligned}
\end{equation}

As in the $R^3$ case, we find that the gauge algebra closes on terms that explicitly depend on the gauge fields. Taking two parameters $X_1,X_2\in E$, we have
\begin{align}
\begin{split}
[ \delta_{X_1} ,\delta_{X_2} ] A &= \mathcal{L}_{[x_1, x_2]} A -\dd \left( [\Lambda_1,\Lambda_2] + i_{x_1} \dd \Lambda_2 -  i_{x_2} \dd \Lambda_1 \right)\\  
& -[A, [\Lambda_1,\Lambda_2] + i_{x_1} \dd \Lambda_2 -  i_{x_2} \dd \Lambda_1 ],\\[5pt]
[\delta_{X_1},\delta_{X_2}] \tilde{C}_6&=  \mathcal{L}_{[x_1, x_2]} \tilde{C}_6 - \dd (i_{x_1} \dd \sigma_2 - i_{x_2} \dd \sigma_1 +\tfrac12 \dd i_{x_1} \sigma_2 -\tfrac12 \dd i_{x_2} \sigma_1) \\  
&   -  \tr\dd \left( [\Lambda_1,\Lambda_2] + i_{x_1} \dd \Lambda_2 -  i_{x_2} \dd \Lambda_1 \right) \big( A (\dd A )^2  \\  
&  \quad + \tfrac25 ( A^3  \dd A +   \dd A A^3  + A^5)+ \tfrac15 ( A^2 \dd A A  +   A \dd A A^2  ) \big)\\  
&  +\dd \tr \big( \Lambda_1 \left(   \dd \Lambda_2 \dd A \dd A + \tfrac35 \dd \Lambda_2 \dd ( A^3) +\tfrac15\dd ( A^2 \dd \Lambda_2 A ) \right) \\ 
& \quad-  \Lambda_2  \left(   \dd \Lambda_1 \dd A \dd A + \tfrac35 \dd \Lambda_1 \dd ( A^3) +\tfrac15 \dd (A^2 \dd \Lambda_1 A )\right)\big) ,
\end{split}
\end{align}
and therefore, the algebra of the gauge parameters is
\begin{equation}\label{eq:7-gauge}
\begin{split}
[X_1,X_2] &= [x_1, x_2] + [\Lambda_1,\Lambda_2] + i_{x_1} \dd \Lambda_2 -  i_{x_2} \dd \Lambda_1 \\
&+ i_{x_1} \dd \sigma_2 - i_{x_2} \dd \sigma_1 +\tfrac12 \dd i_{x_1} \sigma_2 -\tfrac12 \dd i_{x_2} \sigma_1 \\
&- \tr \big( \Lambda_1  \left(   \dd \Lambda_2 \dd A \dd A + \tfrac35 \dd \Lambda_2\dd ( A^3) +\tfrac15 \dd ( A^2 )\dd \Lambda_2 A - \tfrac15 A^2 \dd \Lambda_2 \dd A \right) \\ 
& \quad-  \Lambda_2 \left(   \dd \Lambda_1 \dd A \dd A + \tfrac35 \dd \Lambda_1\dd ( A^3) +\tfrac15 \dd ( A^2 )\dd \Lambda_1 A - \tfrac15 A^2 \dd \Lambda_1 \dd A \right)\big) \in E .
\end{split}
\end{equation}

Let us then see how this fits with the $L_{\infty}$ formalism.

\subsection{An $L_{\infty}$ gauge algebra for $R^4$}

We start by building a seven term graded vector space
\begin{equation}
V = V_5\oplus V_4\oplus V_3\oplus V_2\oplus V_1 \oplus V_0 \oplus V_{-1},
\end{equation}
where
\begin{equation}
\begin{aligned}
& V_5 = C^{\infty}(M), \quad V_4 =    T^*  , \quad V_3 = \Lambda^2 T^*  , \quad V_2 =  \Lambda^3 T^*  , \quad V_1 = \Lambda^4 T^* ,\\[5pt]
  & V_0 = E = T\oplus \mathfrak{g}\oplus  \Lambda^5 T^* ,  \quad   V_{-1}=T^* \otimes\mathfrak{g} \oplus \Lambda^6 T^* ,
\end{aligned}
\end{equation}
whose elements we will generically label as
\begin{equation}
 \xi \in V_1\oplus V_2\oplus V_3\oplus  V_4\oplus V_5, \quad X= x + \Lambda + \sigma \in V_0,\quad  \Psi = A + \tilde{C} \in V_{-1} . 
\end{equation}

We now construct the $L_{\infty}$ products as before.  The terms in the products which are independent of $V_{-1}$ or the $\mathfrak{g}$ part of $V_0$ must reproduce the results of~\cite{zambon}. Again we read off some of the brackets by comparing~\eqref{eq:7-symm} and~\eqref{eq:7-gauge} with~\eqref{eq:Linf-symm} and~\eqref{eq:Linf-gauge} respectively. Then picking a specific point $\Psi$ in the space $V_{-1}$ will correspond to specifying the supergravity data by demanding that the gauge algebra obeyed by $\Psi$ is 
\begin{equation}
\begin{aligned}
\delta_X \Psi  &= \ell_1(X) + \ell_2(X, \Psi) -\tfrac12 \ell_3(X, \Psi^2)-\tfrac16 \ell_4(X,\Psi^3) \\
&\quad+\tfrac{1}{24}\ell_5(X,\Psi^4)+\tfrac{1}{120}\ell_6(X,\Psi^5),
\end{aligned}
\end{equation}
and
\begin{equation}
\begin{split}
[ X_1 , X_2 ]    &= \ell_2(X_1, X_2) + \ell_3(X_1, X_2, \Psi)-\tfrac12 \ell_4(X_1,X_2, \Psi^2) -\tfrac16 \ell_5(X_1,X_2,\Psi^3),
\end{split}
\end{equation}
such that its components $\Psi = A + \tilde{C}_6$ match~\eqref{eq:7-symm} and~\eqref{eq:7-gauge} by construction.

We can then use the generalised Jacobi conditions to complete the algebra. As in the previous section, we can verify the relations by exhaustively going through every term of~\eqref{eq:gen-Jac}, for each level $n$ and for each possible set of inputs for the brackets, since the extra sub-structure of the vector spaces $V_0$ and $V_{-1}$ means that many of those terms vanish trivially and thus the method becomes tractable. The full list of non-vanishing multilinear brackets is nonetheless still rather long. We find the following:

at level one (these show that $V$ is a differential chain complex)
\begin{align}\label{eq:r4-l1}
\ell_1(\xi) = \dd \xi,    \quad \ell_1(X) = -\dd \Lambda - \dd \sigma, \quad \ell_1(\Psi) = 0 ,
\end{align}

at level two (these include the normal gauge transformations)
\begin{subequations}\label{eq:r4-l2}
\begin{align}
\begin{split}
&\ell_2(X,\xi) = \tfrac12 \mathcal{L}_{x}  \xi ,
\end{split}\\[7pt]
\begin{split}
&\ell_2(X_1, X_2) = [x_1, x_2] + [\Lambda_1,\Lambda_2] + \mathcal{L}_{x_1} \Lambda_2 -   \mathcal{L}_{x_2}  \Lambda_1\\&\quad
+  \mathcal{L}_{x_1} \sigma_2 -  \mathcal{L}_{x_2} \sigma_1 -\tfrac12 \dd i_{x_1} \sigma_2 +\tfrac12 \dd i_{x_2} \sigma_1 ,
\end{split}\\[7pt]
\begin{split}
&\ell_2(X,\Psi) = \mathcal{L}_{x}\Psi  - [A , \Lambda] ,
\end{split}
\end{align}
\end{subequations}

at level three (this is the level where the usual Jacobi identity breaks and one is led to use the higher formalism)
\begin{subequations}\label{eq:r4-l3}
\begin{align}
\begin{split}
&\ell_3( \xi, X_1, X_2) = -\tfrac16 ( i_{x_{[1}} \mathcal{L}_{x_{2]}} + i_{[x_1,x_2]} ) \xi  ,
\end{split}\\[7pt]
\begin{split}
&\ell_3(X_1, X_2, X_3) = -\tfrac12\big(i_{x_{[1}} \mathcal{L}_{x_2}+ i_{[x_{[1},x_2]}+ i_{x_{[1}} i_{x_2} \dd \big) \sigma_{3]}   ,
\end{split}
\end{align}
\end{subequations}

 at level four 
\begin{subequations}\label{eq:r4-l4}
\begin{align}
\begin{split}
&\ell_4(X_1, X_2, \Psi_1, \Psi_2) = 4 \tr \Lambda_{[1}\dd \Lambda_{2]} \dd A_{(1}\dd A_{2)} , 
\end{split}\\[7pt]
\begin{split}
&\ell_4(X, \Psi_1, \Psi_2, \Psi_3) =  6 \tr \dd \Lambda A_{(1} \dd A_2 \dd A_{3)}  ,
\end{split}
\end{align}
\end{subequations}

 at level five  
\begin{subequations}\label{eq:r4-l5}
\begin{align}
\begin{split}
&\ell_5( \xi, X_1, X_2, X_3, X_4) = -\tfrac15 i_{x_{[1}} i_{x_2}\, \ell_3( \xi, X_{3}, X_{4]}) ,
\end{split}\\[7pt]
\begin{split}
&\ell_5(   X_1, \dots, X_5) = -\tfrac13 i_{x_{[1}} i_{x_2}\, \ell_3(X_3,X_4,X_{5]}),
\end{split}\\[7pt]
\begin{split}
&\ell_5(   X_1, X_2, X_3, X_4, \Psi) = - \tfrac{24}{5}   \tr \big(2  \Lambda_{[1} \Lambda_2\Lambda_3\dd\Lambda_{4]}    -  \Lambda_{[1}\Lambda_2\dd  \Lambda_3 \Lambda_{4]}      \big)\dd A ,
\end{split}\\[7pt]
\begin{split}
&\ell_5(X_1,X_2,X_3,\Psi_1,\Psi_2) = -\tfrac32 i_{x_{[1}}\,\ell_4(X_2,X_{3]},\Psi_1,\Psi_2) \\  
&\quad-\tfrac{12}{5}  \tr \big(2\Lambda_{[1}\Lambda_2\dd\Lambda_{3]}\dd   A_{(1}  A_{2)}  +3\Lambda_{[1} \Lambda_2\dd \Lambda_{3]} A_{(1} \dd A_{2)} -\Lambda_{[1} \dd  \Lambda_2 \Lambda_{3]} \dd(A_{(1} A_{2)}) \\
&\quad -\Lambda_{[1}\dd\Lambda_2\dd A_{(1}\Lambda_{3]} A_{2)} +\Lambda_{[1}\dd\Lambda_2 A_{(1} \Lambda_{3]} \dd A_{2)}\big) ,
\end{split}\\[7pt]
\begin{split}
&\ell_5(X_1, X_2, \Psi_1, \Psi_2,\Psi_3) =  \tfrac{12}{5} \tr  \big( 3 \Lambda_{[1}\dd \Lambda_{2]}   \dd (A_{(1} A_2 A_{3)}) +  \Lambda_{[1}   \dd ( A_{(1} A_{2 } \dd \Lambda_{2]} A_{3)} ) \big),
\end{split}\\[7pt]
\begin{split}
&\ell_5(X , \Psi_1, \Psi_2, \Psi_3,\Psi_4) = -\tfrac{24}{5} \tr  \dd \Lambda  \big( 2    A_{(1} A_2 A_3  \dd A_{4)}  +  A_{(1} A_2  \dd A_3 A_{4)}  \\  
&\quad+    A_{(1} \dd A_2 A_3 A_{4)}  +  2 \dd A_{(1} A_2 A_3 A_{4)}  \big),
\end{split}
\end{align}
\end{subequations}

 at level six  
\begin{subequations}\label{eq:r4-l6}
\begin{align}
\begin{split}
&\ell_6(X_1,\dots,X_6 )= -144  \tr    \Lambda_{[1} \Lambda_2\Lambda_3\Lambda_4\Lambda_5\dd\Lambda_{6]} ,
\end{split}\\[7pt]
\begin{split}
&\ell_6(X_1,\dots,X_5,\Psi )= 24\tr \big(2  \Lambda_{[1} \Lambda_2\Lambda_3\Lambda_4\dd\Lambda_{5]}    +  \Lambda_{[1}\Lambda_2\dd  \Lambda_3 \Lambda_4\Lambda_{5]}      \big)  A \\
&\quad -\tfrac{5}{2}  i_{x_{[1}} \,\ell_5(   X_2, X_3, X_4, X_{5]}, \Psi),
\end{split}\\[7pt]
\begin{split}
&\ell_6(X_1,X_2,X_3 , X_4,\Psi_1, \Psi_2 )= -\tfrac{48}{5} \tr\big( \Lambda_{[1}\dd(\Lambda_2\Lambda_3)\Lambda_{4]} A_{(1} A_{2)}\\ 
&\quad + \Lambda_{[1}\Lambda_2\dd\Lambda_{3 }A_{(1} \Lambda_{4]} A_{2)} + \Lambda_{[1}\dd\Lambda_2 A_{(1} \Lambda_3 \Lambda_{4]} A_{2)}\big)\\ 
&\quad -2 i_{x_{[1}}\, \ell_5(X_2,X_3,X_{4]},\Psi_1,\Psi_2)  -2 i_{x_{[1}}i_{x_2}\, \ell_4(X_3,X_{4]},\Psi_1,\Psi_2),
\end{split}\\[7pt]
\begin{split}
&\ell_6(X_1,X_2,X_3 , \Psi_1, \Psi_2, \Psi_3 ) =   -\tfrac{3}{2} i_{x_{[1}} \,\ell_5( X_2,X_{3]} , \Psi_1, \Psi_2, \Psi_3 )  \\
&\quad - \tfrac{36}{5} \tr \big(      3  \Lambda_{[1} \Lambda_2 \dd \Lambda_{3]}  A_{(1}A_2  A_{3)}   -\Lambda_{[1} \Lambda_2 A_{(1} A_2\dd \Lambda_{3]} A_{3)}   \big) ,
\end{split}\\[7pt]
\begin{split}
&\ell_6(X , \Psi_1, \dots,\Psi_5) =  - 48 \tr   \dd \Lambda    A_{(1} A_2 A_3 A_4 A_{5)} ,   
\end{split}
\end{align}
\end{subequations}

at level seven (last level with a  bracket acting just on the generalised tangent space $E$, the corresponding higher Courant algebroid of~\cite{zambon} would terminate here)
\begin{subequations}\label{eq:r4-l7}
\begin{align}
\begin{split}
&\ell_7(X_1,\dots,X_7) = -144   \tr   \Lambda_{[1} \Lambda_2\Lambda_3\Lambda_4\Lambda_5 \Lambda_6\Lambda_{7]} -\tfrac{7}{2}   i_{x_{[1}}\, \ell_6( X_2,\dots,X_{7]})   \\ 
&\quad +\tfrac13  i_{x_{[1}} i_{x_2}i_{x_3}i_{x_4}\,\ell_3(X_5,X_6,X_{7]}),
\end{split}\\[7pt]
\begin{split}
&\ell_7(X_1,\dots,X_6,\Psi  ) = - 3  i_{x_{[1}}\,\ell_6( X_2,\dots,X_{6]},\Psi  )    -5   i_{x_{[1}}i_{x_2}\, \ell_5( X_3,X_4,X_5,X_{6]},\Psi  ),
\end{split}\\[7pt]
\begin{split}
&\ell_7(X_1,\dots,X_5,\Psi_1,\Psi_2 ) =  -\tfrac52   i_{x_{[1}}\ell_6( X_2,X_3,X_4,X_{5]},\Psi_1,\Psi_2 )\\ 
&\quad -\tfrac{10}{3}  i_{x_{[1}}i_{x_2} \,\ell_5( X_3,X_4,X_{5]},\Psi_1,\Psi_2 )    -\tfrac52   i_{x_{[1}}i_{x_2}i_{x_3}\, \ell_4 ( X_4,X_{5]},\Psi_1,\Psi_2 )  ,
\end{split}\\[7pt]
\begin{split}
&\ell_7(X_1,X_2,X_3,X_4,\Psi_1,\Psi_2,\Psi_3) =  -2 i_{x_{[1}}\,\ell_6( X_2,X_3,X_{4]},\Psi_1,\Psi_2,\Psi_3) \\ 
&\quad  -2 i_{x_{[1}}i_{x_2} \,\ell_5( X_3,X_{4]},\Psi_1,\Psi_2,\Psi_3),
\end{split}
\end{align}
\end{subequations}

 at level eight
\begin{subequations}\label{eq:r4-l8}
\begin{align}
\begin{split}
&\ell_8(X_1,\dots,X_7 ,\Psi   )= \tfrac{7}{2} i_{x_{[1}}   i_{x_{2}}\, \ell_6(X_3,\dots,X_{7]} ,\Psi   ) \\&\quad+\tfrac{35}{4}  i_{x_{[1}}   i_{x_{2}} i_{x_3}\,\ell_5( X_4,X_5,X_6,X_{7]} ,\Psi   ),
\end{split}\\[7pt]
\begin{split}
&\ell_8(X_1,\dots,X_6,\Psi_1,\Psi_2 ) =  \tfrac52  i_{x_{[1}}i_{x_2}\,\ell_6( X_4,X_5,X_{6]},\Psi_1,\Psi_2 )  \\ 
&\quad   +5 i_{x_{[1}}i_{x_2} i_{x_3} \,\ell_5(X_4,X_5,X_{6]},\Psi_1,\Psi_2) + \tfrac92  i_{x_{[1}}i_{x_2} i_{x_3}i_{x_4}\,\ell_4(X_5,X_{6]},\Psi_1,\Psi_2),
\end{split}\\[7pt]
\begin{split}
&\ell_8(X_1,\dots,X_5,\Psi_1,\Psi_2,\Psi_3) = \tfrac53 i_{x_{[1}}i_{x_2}\, \ell_6(X_3 ,X_4,X_{5]},\Psi_1,\Psi_2,\Psi_3) \\ 
&\quad  +\tfrac52 i_{x_{[1}}i_{x_2} i_{x_3} \, \ell_5( X_4,X_{5]},\Psi_1,\Psi_2,\Psi_3),
\end{split}
\end{align}
\end{subequations}

 at level nine
\begin{subequations}\label{eq:r4-l9}
\begin{align}
\begin{split}
&\ell_9(X_1,\dots,X_7,\Psi_1,\Psi_2)   = -\tfrac{7}{6}  i_{x_{[1}} i_{x_2}i_{x_3}i_{x_4} \,\ell_5( X_5,X_6,X_{7]},\Psi_1,\Psi_2)
 \\ 
 &\quad  -\tfrac{7}{4}  i_{x_{[1}} i_{x_2}i_{x_3}i_{x_4}i_{x_5} \,\ell_4(  X_6,X_{7]},\Psi_1,\Psi_2) ,
\end{split}\\[7pt]
\begin{split}
&\ell_9(X_1,\dots,X_6,\Psi_1,\Psi_2,\Psi_3)  = -\tfrac12 i_{x_{[1}}i_{x_2}i_{x_3}i_{x_4}\,\ell_5( X_5,X_{6]},\Psi_1,\Psi_2,\Psi_3)  ,
\end{split}
\end{align}
\end{subequations}

and finally at level ten
\begin{align}\label{eq:r4-l10}
\begin{split}
&\ell_{10}(X_1,\dots,X_7,\Psi_1,\Psi_2,\Psi_3)  = - \tfrac{7}{6}  i_{x_{[1}} i_{x_2}i_{x_3}i_{x_4}\,\ell_6( X_5,X_6,X_{7]},\Psi_1,\Psi_2,\Psi_3) \\
&\quad- \tfrac{7}{4}     i_{x_{[1}} i_{x_2}i_{x_3}i_{x_4}i_{x_5}\,\ell_5( X_6,X_{7]},\Psi_1,\Psi_2,\Psi_3) .
\end{split}
\end{align}

All other brackets vanish. As in the previous section, we observe that most of terms in the brackets can be expressed recursively, which is to be expected since they were built by explicitly iterating through the generalised Jacobi identities.  Note also that, as mentioned earlier, the terms that depend only on elements  $\Lambda$ and $A$  are simply reproducing the (polarised)  $p$-forms that result from the descent equations of the anomaly polynomial. For example, we have that 
\begin{equation}
\omega^1_6(\Lambda,A) = \tfrac{1}{3!}\ell_4(\Lambda,A^3)-\tfrac{1}{4!}\ell_5(\Lambda,A^4)-\tfrac{1}{5!}\ell_6(\Lambda,A^5).
\end{equation}

Despite no longer being able to describe these higher order gauge algebras in terms of just a Leibniz bracket on the generalised tangent space, we thus have that the extra structure of $E$ is still enough to ensure that we can find an $L_{\infty}$ algebra, and that this algebra has a finite number of brackets. And while there should not be much difficulty in adding the extra geometrical data that make up the physical degrees of freedom such as the Riemannian metric, it will require further study to see whether this weaker differential structure will be enough to give a natural geometric description of the dynamics of higher-derivative-corrected supergravity.

\begin{acknowledgments}
I would like to thank Ruben Minasian and Dan Waldram for helpful discussions. 
This work has been supported by the European Research Council (ERC) under the European Union's Horizon 2020 research and innovation programme (``Exceptional Quantum Gravity'', grant agreement No 740209).
\end{acknowledgments}

\appendix*
\section{Conventions and Chern-Simons forms}
\label{app:def}
We mostly follow the conventions of~\cite{het-gg}, though we generally omit the wedge symbol for the product of differential forms.

Given a Lie algebra-valued one-form $A$ we define its curvature by 
\begin{equation}
R(A) = \dd A + A^2,
\end{equation}
which satisfies
\begin{equation}
\dd_A R = \dd R + [A,R] = 0 ,
\end{equation}
and is invariant under the infinitesimal gauge transformations of the potential
\begin{equation}
\delta_\Lambda A = - \dd_A \Lambda = - \dd\Lambda - [A,\Lambda] .
\end{equation}

As is well known from the study of anomalies~\cite{Zumino:1983ew, zumino-zee, AlvarezGaume:1984dr, Manes:1985df}, taking the trace of powers of the curvature one can define invariant polynomials
\begin{equation}
\dd \tr R^n = \delta \tr R^n = 0 ,
\end{equation}
from the $n$-th Chern character $\tr R^n$ of the gauge vector bundle. Poincar\'e's lemma then implies that one can locally define the Chern-Simons forms $\omega_{(2n-1)}$
\begin{equation}
\dd \omega_{(2n-1)}(A) = \tr R^n ,
\end{equation}
and applying the lemma once again, now for $\delta$, gives
\begin{equation}
\dd \omega^1_{(2n-2)}(\Lambda, A) = \delta_{\Lambda} \omega_{(2n-1)}(A) ,
\end{equation}
where the superscript denotes the powers of the gauge parameter, since, in principle, one can continue ``descending'' along this chain.

We can list (up to exact terms) some of the Chern-Simons forms that will be important for us explicitly
\begin{align}
\omega_7(A) &= \tr \left( A (\dd A )^3 + \tfrac85 A^3  (\dd A)^2 + \tfrac45 A \dd A A^2 \dd A + 2 A^5 \dd A +\tfrac47 A^7 \right),\\[5pt]
\begin{split}\label{eq:app-omega6}
\omega^1_6(\Lambda,A) &=  \tr  \dd \Lambda\big( A (\dd A )^2 + \tfrac25 ( A^3  \dd A +   \dd A A^3  + A^5)\\ 
&\quad + \tfrac15 ( A^2 \dd A A  +   A \dd A A^2  ) \big) ,\end{split}\\[5pt]
\omega_5(A) &= \tr \left( A (\dd A )^2 + \tfrac32 A^3  \dd A + \tfrac35 A^5 \right),\\[5pt]
\omega^1_4(\Lambda,A) &=  \tr  \dd \Lambda \left( A  \dd A  + \tfrac12   A^3 \right) ,\\[5pt]
\omega_3(A) &= \tr \left( A \dd A + \tfrac23 A^3\right) , \\[5pt]
 \omega^1_2(\Lambda,A)&= \tr\dd\Lambda A .
\end{align}
The last two are not used in this work but are the ones that are featured in Heterotic generalised geometry. These agree with the usual ones in the literature~\cite{zumino-zee} up to exact terms corresponding to our convention choice of having the differential acting on the parameter $\Lambda$.


\begin{thebibliography}{99}

\bibitem{Hitchin:2004ut}
  N.~Hitchin,
  ``Generalized Calabi-Yau manifolds,''
  Quart.\ J.\ Math.\  {\bf 54} (2003) 281
  doi:10.1093/qjmath/54.3.281
  [math/0209099 [math-dg]].
  
\bibitem{gualtieri}
   M.~Gualtieri, 
   ``Generalized Complex Geometry,''
   Oxford University DPhil thesis (2004) [arXiv:math.DG/0401221] 
   and [arXiv:math.DG/0703298].
   
\bibitem{dorfman}  
  I.~Dorfman,
  ``Dirac structures of integrable evolution equations,''
  Phys.\ Lett.\ A, {\bf 125} (1987) 240
  doi:10.1016/0375-9601(87)90201-5 
     
\bibitem{courant}
	T.~Courant, 
	``Dirac manifolds,''
	Trans.\  Amer.\  Math.\ Soc.\ {\bf 319} (1990) 631 
	doi:10.2307/2001258
	
\bibitem{Liu:1995lsa}
  Z.~J.~Liu, A.~Weinstein and P.~Xu,
  ``Manin Triples for Lie Bialgebroids,''
  J.\ Diff.\ Geom.\  {\bf 45} (1997) no.3,  547
  [dg-ga/9508013].
  
\bibitem{severa}
	P.~Severa,
	Letter to Alan Weinstein. 
	http://sophia.dtp.fmph.uniba.sk/~severa/letters/
  
\bibitem{Hitchin:2005in}
  N.~Hitchin,
  ``Brackets, forms and invariant functionals,''
  math/0508618 [math-dg].
  
\bibitem{Grana:2004bg}
  M.~Grana, R.~Minasian, M.~Petrini and A.~Tomasiello,
  ``Supersymmetric backgrounds from generalized Calabi-Yau manifolds,''
  JHEP {\bf 0408} (2004) 046
  doi:10.1088/1126-6708/2004/08/046
  [hep-th/0406137].

\bibitem{Grana:2005sn}
  M.~Grana, R.~Minasian, M.~Petrini and A.~Tomasiello,
  ``Generalized structures of N=1 vacua,''
  JHEP {\bf 0511} (2005) 020
  doi:10.1088/1126-6708/2005/11/020
  [hep-th/0505212].
  
\bibitem{Lindstrom:2004eh}
  U.~Lindstrom,
  ``Generalized N = (2,2) supersymmetric nonlinear sigma models,''
  Phys.\ Lett.\ B {\bf 587} (2004) 216
  doi:10.1016/j.physletb.2004.03.014
  [hep-th/0401100].  
  
\bibitem{Lindstrom:2004iw}
  U.~Lindstrom, R.~Minasian, A.~Tomasiello and M.~Zabzine,
  ``Generalized complex manifolds and supersymmetry,''
  Commun.\ Math.\ Phys.\  {\bf 257} (2005) 235
  doi:10.1007/s00220-004-1265-6
  [hep-th/0405085].  

\bibitem{Coimbra:2011nw}
  A.~Coimbra, C.~Strickland-Constable and D.~Waldram,
  ``Supergravity as Generalised Geometry I: Type II Theories,''
  JHEP {\bf 1111} (2011) 091
  doi:10.1007/JHEP11(2011)091
  [arXiv:1107.1733 [hep-th]].

\bibitem{Aldazabal:2013sca}
  G.~Aldazabal, D.~Marques and C.~Nunez,
  ``Double Field Theory: A Pedagogical Review,''
  Class.\ Quant.\ Grav.\  {\bf 30} (2013) 163001
  doi:10.1088/0264-9381/30/16/163001
  [arXiv:1305.1907 [hep-th]].
  
\bibitem{Green:1984sg}
  M.~B.~Green and J.~H.~Schwarz,
  ``Anomaly Cancellation in Supersymmetric D=10 Gauge Theory and Superstring Theory,''
  Phys.\ Lett.\  {\bf 149B} (1984) 117.
  doi:10.1016/0370-2693(84)91565-X  
  
\bibitem{Garcia-Fernandez:2013gja}
  M.~Garcia-Fernandez,
   ``Torsion-free generalized connections and Heterotic Supergravity,''
  Commun.\ Math.\ Phys.\  {\bf 332} (2014) no.1,  89
  doi:10.1007/s00220-014-2143-5
  [arXiv:1304.4294 [math.DG]].  
  
\bibitem{Baraglia:2013wua}
  D.~Baraglia and P.~Hekmati,
   ``Transitive Courant Algebroids, String Structures and T-duality,''
  Adv.\ Theor.\ Math.\ Phys.\  {\bf 19} (2015) 613
  doi:10.4310/ATMP.2015.v19.n3.a3
  [arXiv:1308.5159 [math.DG]].  
   
\bibitem{Anderson:2014xha}
  L.~B.~Anderson, J.~Gray and E.~Sharpe,
   ``Algebroids, Heterotic Moduli Spaces and the Strominger System,''
  JHEP {\bf 1407} (2014) 037
  doi:10.1007/JHEP07(2014)037
  [arXiv:1402.1532 [hep-th]].
  
\bibitem{delaOssa:2014cia}
  X.~de la Ossa and E.~E.~Svanes,
   ``Holomorphic Bundles and the Moduli Space of N=1 Supersymmetric Heterotic Compactifications,''
  JHEP {\bf 1410} (2014) 123
  doi:10.1007/JHEP10(2014)123
  [arXiv:1402.1725 [hep-th]].
    
\bibitem{het-gg}
  A.~Coimbra, R.~Minasian, H.~Triendl and D.~Waldram,
  ``Generalised geometry for string corrections,''
  JHEP {\bf 1411} (2014) 160
  doi:10.1007/JHEP11(2014)160
  [arXiv:1407.7542 [hep-th]].
  
\bibitem{Bedoya:2014pma}
  O.~A.~Bedoya, D.~Marques and C.~Nunez,
   ``Heterotic $\alpha$'-corrections in Double Field Theory,''
  JHEP {\bf 1412} (2014) 074
  doi:10.1007/JHEP12(2014)074
  [arXiv:1407.0365 [hep-th]].  
   
\bibitem{Witten:1999eg}
  E.~Witten,
  ``World sheet corrections via D instantons,''
  JHEP {\bf 0002} (2000) 030
  doi:10.1088/1126-6708/2000/02/030
  [hep-th/9907041].
 
\bibitem{Bergshoeff:1988nn}
  E.~Bergshoeff and M.~de Roo,
  ``Supersymmetric Chern-simons Terms in Ten-dimensions,''
  Phys.\ Lett.\ B {\bf 218} (1989) 210.
  doi:10.1016/0370-2693(89)91420-2  
  
\bibitem{Cremmer:1978km}
  E.~Cremmer, B.~Julia and J.~Scherk,
  ``Supergravity Theory in Eleven-Dimensions,''
  Phys.\ Lett.\ B {\bf 76} (1978) 409
   [Phys.\ Lett.\  {\bf 76B} (1978) 409].
  doi:10.1016/0370-2693(78)90894-8  
  
\bibitem{Vafa:1995fj}
  C.~Vafa and E.~Witten,
 ``A One loop test of string duality,''
  Nucl.\ Phys.\ B {\bf 447} (1995) 261
  doi:10.1016/0550-3213(95)00280-6
  [hep-th/9505053].
  
\bibitem{Duff:1995wd}
  M.~J.~Duff, J.~T.~Liu and R.~Minasian,
  ``Eleven-dimensional origin of string-string duality: A One loop test,''
  Nucl.\ Phys.\ B {\bf 452} (1995) 261
  doi:10.1016/0550-3213(95)00368-3
  [hep-th/9506126].   
  
\bibitem{Coimbra:2012af}
  A.~Coimbra, C.~Strickland-Constable and D.~Waldram,
  ``Supergravity as Generalised Geometry II: $E_{d(d)} \times \mathbb{R}^+$ and M theory,''
  JHEP {\bf 1403} (2014) 019
  doi:10.1007/JHEP03(2014)019
  [arXiv:1212.1586 [hep-th]].
  
\bibitem{Hull:2007zu}
  C.~M.~Hull,
  ``Generalised Geometry for M-Theory,''
  JHEP {\bf 0707} (2007) 079
  doi:10.1088/1126-6708/2007/07/079
  [hep-th/0701203].
  
\bibitem{Pacheco:2008ps}
  P.~Pires Pacheco and D.~Waldram,
  ``M-theory, exceptional generalised geometry and superpotentials,''
  JHEP {\bf 0809} (2008) 123
  doi:10.1088/1126-6708/2008/09/123
  [arXiv:0804.1362 [hep-th]].

\bibitem{e7gg}
  A.~Coimbra, C.~Strickland-Constable and D.~Waldram,
  ``$E_{d(d)} \times \mathbb{R}^+$ generalised geometry, connections and M theory,''
  JHEP {\bf 1402} (2014) 054
  doi:10.1007/JHEP02(2014)054
  [arXiv:1112.3989 [hep-th]]. 
   
\bibitem{Hohm:2014fxa}
  O.~Hohm and H.~Samtleben,
  ``Exceptional field theory. III. E$_{8(8)}$,''
  Phys.\ Rev.\ D {\bf 90} (2014) 066002
  doi:10.1103/PhysRevD.90.066002
  [arXiv:1406.3348 [hep-th]].
   
\bibitem{deWit:1986mz}
  B.~de Wit and H.~Nicolai,
  ``$d=11$ Supergravity With Local SU(8) Invariance,''
  Nucl.\ Phys.\ B {\bf 274} (1986) 363.
  doi:10.1016/0550-3213(86)90290-7
  
\bibitem{Grana:2011nb}
  M.~Grana and F.~Orsi,
  ``N=1 vacua in Exceptional Generalized Geometry,''
  JHEP {\bf 1108} (2011) 109
  doi:10.1007/JHEP08(2011)109
  [arXiv:1105.4855 [hep-th]].  
  
\bibitem{Grana:2012ea}
  M.~Grana and F.~Orsi,
  ``N=2 vacua in Generalized Geometry,''
  JHEP {\bf 1211} (2012) 052
  doi:10.1007/JHEP11(2012)052
  [arXiv:1207.3004 [hep-th]].
  
\bibitem{Coimbra:2014uxa}
  A.~Coimbra, C.~Strickland-Constable and D.~Waldram,
  ``Supersymmetric Backgrounds and Generalised Special Holonomy,''
  Class.\ Quant.\ Grav.\  {\bf 33} (2016) no.12,  125026
  doi:10.1088/0264-9381/33/12/125026
  [arXiv:1411.5721 [hep-th]].  
  
\bibitem{Ashmore:2015joa}
  A.~Ashmore and D.~Waldram,
   ``Exceptional Calabi-Yau spaces: the geometry of $\mathcal{N}=2$ backgrounds with flux,''
  Fortsch.\ Phys.\  {\bf 65} (2017) no.1,  1600109
  doi:10.1002/prop.201600109
  [arXiv:1510.00022 [hep-th]].  
  
\bibitem{Coimbra:2015nha}
  A.~Coimbra and C.~Strickland-Constable,
  ``Generalised Structures for $\mathcal{N}=1$ AdS Backgrounds,''
  JHEP {\bf 1611} (2016) 092
  doi:10.1007/JHEP11(2016)092
  [arXiv:1504.02465 [hep-th]].  
  
\bibitem{Ashmore:2016qvs}
  A.~Ashmore, M.~Petrini and D.~Waldram,
  ``The exceptional generalised geometry of supersymmetric AdS flux backgrounds,''
  JHEP {\bf 1612} (2016) 146
  doi:10.1007/JHEP12(2016)146
  [arXiv:1602.02158 [hep-th]].     

\bibitem{Coimbra:2016ydd}
  A.~Coimbra and C.~Strickland-Constable,
   ``Supersymmetric Backgrounds, the Killing Superalgebra, and Generalised Special Holonomy,''
  JHEP {\bf 1611} (2016) 063
  doi:10.1007/JHEP11(2016)063
  [arXiv:1606.09304 [hep-th]].
  
\bibitem{Malek:2017njj}
  E.~Malek,
  ``Half‐Maximal Supersymmetry from Exceptional Field Theory,''
  Fortsch.\ Phys.\  {\bf 65} (2017) no.10-11,  1700061
  doi:10.1002/prop.201700061
  [arXiv:1707.00714 [hep-th]].
  
\bibitem{Coimbra:2017fqv}
  A.~Coimbra and C.~Strickland-Constable,
   ``Supersymmetric AdS backgrounds and weak generalised holonomy,''
  arXiv:1710.04156 [hep-th].
  
\bibitem{Ashmore:2016oug}
  A.~Ashmore, M.~Gabella, M.~Graña, M.~Petrini and D.~Waldram,
  ``Exactly marginal deformations from exceptional generalised geometry,''
  JHEP {\bf 1701} (2017) 124
  doi:10.1007/JHEP01(2017)124
  [arXiv:1605.05730 [hep-th]].

\bibitem{Ashmore:2018npi}
  A.~Ashmore,
   ``Marginal deformations of 3d $\mathcal{N}=2$ CFTs from AdS$_4$ backgrounds in generalised geometry,''
  JHEP {\bf 1812} (2018) 060
  doi:10.1007/JHEP12(2018)060
  [arXiv:1809.03503 [hep-th]].

\bibitem{Coimbra:2017fqj}
  A.~Coimbra and R.~Minasian,
   ``M-theoretic Lichnerowicz formula and supersymmetry,''
   JHEP {\bf 1910} (2019) 036
   doi:10.1007/JHEP10(2019)036
   [arXiv:1705.04330 [hep-th]].
  
\bibitem{hohm-zwiebach}
  O.~Hohm and B.~Zwiebach,
  ``$L_{\infty}$ Algebras and Field Theory,''
  Fortsch.\ Phys.\  {\bf 65} (2017) no.3-4,  1700014
  doi:10.1002/prop.201700014
  [arXiv:1701.08824 [hep-th]].
  
\bibitem{zambon}
  M.~Zambon,
  ``L-infinity algebras and higher analogues of Dirac structures and Courant algebroids,''
  J.\ Symplectic Geom.\  {\bf 10} (2012) 563
  doi:10.4310/JSG.2012.v10.n4.a4
  [arXiv:1003.1004 [math.SG]].
  
\bibitem{Manes:1985df}
  J.~Manes, R.~Stora and B.~Zumino,
  ``Algebraic Study of Chiral Anomalies,''
  Commun.\ Math.\ Phys.\  {\bf 102} (1985) 157.
  doi:10.1007/BF01208825  
  
\bibitem{Zwiebach:1992ie}
  B.~Zwiebach,
  ``Closed string field theory: Quantum action and the B-V master equation,''
  Nucl.\ Phys.\ B {\bf 390} (1993) 33
  doi:10.1016/0550-3213(93)90388-6
  [hep-th/9206084].  
  
\bibitem{Lada:1992wc}
  T.~Lada and J.~Stasheff,
  ``Introduction to SH Lie algebras for physicists,''
  Int.\ J.\ Theor.\ Phys.\  {\bf 32} (1993) 1087
  doi:10.1007/BF00671791
  [hep-th/9209099].
  
\bibitem{Stasheff:2018vnl}
  J.~Stasheff,
  ``$L_\infty$ and $A_\infty$ structures: then and now,''
  [arXiv:1809.02526 [math.QA]].
  
\bibitem{roytenberg-weinstein}
  D.~Roytenberg and A.~Weinstein,
  ``Courant Algebroids and Strongly Homotopy Lie Algebras,''
  [math/9802118 [math.QA]].

\bibitem{Ashmore:2018ybe}
  A.~Ashmore, X.~De La Ossa, R.~Minasian, C.~Strickland-Constable and E.~E.~Svanes,
  ``Finite deformations from a heterotic superpotential: holomorphic Chern-Simons and an $L_\infty$ algebra,''
  JHEP {\bf 1810} (2018) 179
  doi:10.1007/JHEP10(2018)179
  [arXiv:1806.08367 [hep-th]].
  
\bibitem{Baraglia:2011dg}
  D.~Baraglia,
  ``Leibniz algebroids, twistings and exceptional generalized geometry,''
  J.\ Geom.\ Phys.\  {\bf 62} (2012) 903
  doi:10.1016/j.geomphys.2012.01.007
  [arXiv:1101.0856 [math.DG]].  
  
\bibitem{Arvanitakis:2018cyo}
  A.~S.~Arvanitakis,
  ``Brane Wess-Zumino terms from AKSZ and exceptional generalised geometry as an $L_\infty$-algebroid,''
  arXiv:1804.07303 [hep-th].  
  
\bibitem{Arvanitakis:2019cxy}
  A.~S.~Arvanitakis,
  ``Generalising Courant algebroids to M-theory,''
  PoS CORFU {\bf 2018} (2019) 127
  doi:10.22323/1.347.0127
  [arXiv:1904.12361 [math-ph]].
  
\bibitem{Fiorenza:2019ckz}
  D.~Fiorenza, H.~Sati and U.~Schreiber,
  ``The Rational Higher Structure of M-theory,''
  Fortschritte der Physik, May 2019
  doi:10.1002/prop.201910017
  [arXiv:1903.02834 [hep-th]].

\bibitem{Sati:2009ic}
  H.~Sati, U.~Schreiber and J.~Stasheff,
  ``Differential twisted String and Fivebrane structures,''
  Commun.\ Math.\ Phys.\  {\bf 315} (2012) 169
  doi:10.1007/s00220-012-1510-3
  [arXiv:0910.4001 [math.AT]].
  
\bibitem{Fiorenza:2019usl}
  D.~Fiorenza, H.~Sati and U.~Schreiber,
  ``Twisted Cohomotopy implies M-Theory anomaly cancellation,''
  arXiv:1904.10207 [hep-th].

\bibitem{Deser:2014mxa}
  A.~Deser and J.~Stasheff,
  ``Even symplectic supermanifolds and double field theory,''
  Commun.\ Math.\ Phys.\  {\bf 339} (2015) no.3,  1003
  doi:10.1007/s00220-015-2443-4
  [arXiv:1406.3601 [math-ph]].
  
\bibitem{Wang:2015hca}
  Y.~N.~Wang,
  ``Generalized Cartan Calculus in general dimension,''
  JHEP {\bf 1507} (2015) 114
  doi:10.1007/JHEP07(2015)114
  [arXiv:1504.04780 [hep-th]].
  
\bibitem{Deser:2016qkw}
  A.~Deser and C.~Sämann,
  ``Extended Riemannian Geometry I: Local Double Field Theory,''
  C. Ann. Henri Poincaré (2018) 19: 2297
  doi:10.1007/s00023-018-0694-2
  [arXiv:1611.02772 [hep-th]].
  
\bibitem{Lavau:2017tvi}
  S.~Lavau,
  ``Tensor hierarchies and Leibniz algebras,''
  J.\ Geom.\ Phys.\  {\bf 144} (2019) 147
  doi:10.1016/j.geomphys.2019.05.014
  [arXiv:1708.07068 [hep-th]].  
  
\bibitem{Hohm:2017cey}
  O.~Hohm, V.~Kupriyanov, D.~Lust and M.~Traube,
  ``Constructions of $L_{\infty}$ algebras and their field theory realizations,''
  Adv.\ Math.\ Phys.\  {\bf 2018} (2018) 9282905
  doi:10.1155/2018/9282905
  [arXiv:1709.10004 [math-ph]].  
  
\bibitem{Deser:2017fko}
  A.~Deser, M.~A.~Heller and C.~Sämann,
  ``Extended Riemannian Geometry II: Local Heterotic Double Field Theory,''
  JHEP {\bf 1804} (2018) 106
  doi:10.1007/JHEP04(2018)106
  [arXiv:1711.03308 [hep-th]].
  
\bibitem{Cederwall:2018aab}
  M.~Cederwall and J.~Palmkvist,
  ``$L_\infty$ algebras for extended geometry from Borcherds superalgebras,''
  Commun.\ Math.\ Phys.\  {\bf 369} (2019) no.2,  721
  doi:10.1007/s00220-019-03451-2
  [arXiv:1804.04377 [hep-th]].  
  
\bibitem{Hohm:2018ybo}
  O.~Hohm and H.~Samtleben,
   ``Leibniz-Chern-Simons Theory and Phases of Exceptional Field Theory,''
  Commun.\ Math.\ Phys.\  (2019) 1
  doi:10.1007/s00220-019-03347-1
  [arXiv:1805.03220 [hep-th]].  
  
\bibitem{Cagnacci:2018buk}
  Y.~Cagnacci, T.~Codina and D.~Marques,
  ``$L_\infty$ algebras and Tensor Hierarchies in Exceptional Field Theory and Gauged Supergravity,''
  JHEP {\bf 1901} (2019) 117
  doi:10.1007/JHEP01(2019)117
  [arXiv:1807.06028 [hep-th]].
  
\bibitem{Deser:2018flj}
  A.~Deser and C.~Sämann,
  ``Extended Riemannian Geometry III: Global Double Field Theory with Nilmanifolds,''
  JHEP {\bf 1905} (2019) 209
  doi:10.1007/JHEP05(2019)209
  [arXiv:1812.00026 [hep-th]].
  
\bibitem{Hohm:2019wql}
  O.~Hohm and H.~Samtleben,
  ``Higher Gauge Structures in Double and Exceptional Field Theory,''
  Fortsch.\ Phys.\  {\bf 67} (2019) no.8-9,  1910008
  doi:10.1002/prop.201910008
  [arXiv:1903.02821 [hep-th]].
  
\bibitem{Bonezzi:2019ygf}
  R.~Bonezzi and O.~Hohm,
  ``Leibniz Gauge Theories and Infinity Structures,''
  arXiv:1904.11036 [hep-th].
  
\bibitem{Lavau:2019oja}
  S.~Lavau and J.~Palmkvist,
  ``Infinity-enhancing Leibniz algebras,''
  arXiv:1907.05752 [hep-th].
  
\bibitem{Jurco:2018sby}
  B.~Jurčo, L.~Raspollini, C.~Sämann and M.~Wolf,
  ``$L_\infty$-Algebras of Classical Field Theories and the Batalin-Vilkovisky Formalism,''
  Fortsch.\ Phys.\  {\bf 67} (2019) no.7,  1900025
  doi:10.1002/prop.201900025
  [arXiv:1809.09899 [hep-th]].
  
\bibitem{Strickland-Constable:2013xta}
  C.~Strickland-Constable,
  ``Subsectors, Dynkin Diagrams and New Generalised Geometries,''
  JHEP {\bf 1708} (2017) 144
  doi:10.1007/JHEP08(2017)144
  [arXiv:1310.4196 [hep-th]].

\bibitem{Zumino:1983ew}
  B.~Zumino,
  ``Chiral Anomalies And Differential Geometry: Lectures Given At Les Houches, August 1983,''
  In Treiman, S.b. ( Ed.) Et Al.: Current Algebra and Anomalies, 361-391, World Publishing, Singapore
  
\bibitem{zumino-zee}
  B.~Zumino, Y.~S.~Wu and A.~Zee,
  ``Chiral Anomalies, Higher Dimensions, and Differential Geometry,''
  Nucl.\ Phys.\ B {\bf 239} (1984) 477.
  doi:10.1016/0550-3213(84)90259-1

\bibitem{AlvarezGaume:1984dr}
  L.~Alvarez-Gaume and P.~H.~Ginsparg,
  ``The Structure of Gauge and Gravitational Anomalies,''
  Annals Phys.\  {\bf 161} (1985) 423
   Erratum: [Annals Phys.\  {\bf 171} (1986) 233].
  doi:10.1016/0003-4916(85)90087-9

\end{thebibliography}
\end{document}